\documentclass[a4paper,11pt]{article}
\usepackage{pos}

\begin{document}

\title{Study of the jet transport coefficient at the Large Hadron Collider energies using Color String Percolation Model}
\ShortTitle{}

\author*[a]{Dushmanta Sahu}
\author[b]{Aditya Nath Mishra}
\author[a,c]{Raghunath Sahoo}

\affiliation[a]{Department of Physics, Indian Institute of Technology Indore, 
Simrol, Indore 453552, India}
\affiliation[b]{Wigner Research Centre for Physics, 
  29-33 Konkoly-Thege Miklos str., 1121 Budapest, Hungary}
\affiliation[c]{CERN, CH 1211, Geneva 23, Switzerland}

\emailAdd{Aditya.Nath.Mishra@cern.ch}
\emailAdd{Dushmanta.Sahu@cern.ch}
\emailAdd{Raghunath.Sahoo@cern.ch}

\abstract{In order to have a better understanding of the matter formed in ultra-relativistic collisions, we estimate the jet transport coefficient, $\hat q$, within the Color Sting Percolation Model (CSPM) for various multiplicity classes in proton-proton collisions and centrality classes in nucleus-nucleus collisions at the Large Hadron Collider energies. We study $\hat q$ as a function of final state charged-particle multiplicity and initial percolation temperature. Finally, we compare our obtained results with those calculated from the JET collaboration.}

\FullConference{%
 
 The Ninth Annual Conference on Large Hadron Collider Physics - LHCP2021\\
 7-12 June 2021\\
 Online
}


\maketitle
 
 \section{Introduction}
 
 Jets are collimated emissions of a multitude of hadrons originating from the hard partonic scattering during high-energy collisions. They play a significant role as hard probes of Quark-Gluon Plasma (QGP). These jets lose energy through collisional energy loss and medium-induced gluon radiation \cite{Blaizot:1986}. As a consequence, one can observe the suppression of high transverse momentum particles. This phenomenon is known as jet quenching, which is considered as a direct signature of QGP. At the Relativistic Heavy Ion Collider (RHIC), the evidence of jet quenching has been observed via the measurement of inclusive hadron and jet production at high $p_{\rm T}$, $\gamma$-hadron correlation, di-hadron angular correlations, and the dijet energy imbalance \cite{Adcox:2001jp,Adler:2002tq}. In addition, several theoretical models take parton energy loss into account to study the jet quenching effects, such as; Gyulassy-Levai-Vitev (GLV) and its CUJET implementation, high-twist approach (HT-M and HT BW), Arnold-Moore-Yaffe (AMY) model, etc. In our work \cite{Nath:2021mlp}, we use the QCD-inspired Color String Percolation Model (CSPM) to estimate the jet transport coefficient for various collision species at the LHC energies. The jet transport coefficient describes the average transverse momentum squared transferred from the traversing parton, per unit mean free path. It is widely used to quantify the energy loss of jets in a strongly interacting QCD medium.
 
\section{Formulation}

CSPM is an established theoretical model, which assumes that color strings are stretched between the partons of the colliding nuclei, and the hadronization of these strings produces the final state particles \cite{Phyreport}. These strings occupy a finite area in the transverse space. With the increase in collision energy, the number of colliding partons increases, thus the number of strings also grows and they start overlapping, forming clusters in the transverse space. After a certain critical string density ($\xi_{c}$), a macroscopic cluster appears, which corresponds to the percolation phase transition. In 2D percolation theory, the dimensionless string density parameter is expressed as, $\xi = \frac{N_{s}S_{1}}{S_{N}}$, where $N_{s}$ is the number of overlapping strings, $S_{1}$ is the transverse area of a single string and $S_{N}$ being the transverse area of occupied by the overlapping strings. In the thermodynamic limit, the color suppression factor, $F(\xi)$ is related to the percolation density parameter $\xi $ as \cite{Phyreport},

\begin{eqnarray}
	F(\xi) =\sqrt\frac{1-e^{-\xi}}{\xi}.
	\label{eq1}
\end{eqnarray}

$F(\xi)$ is the color suppression factor by which the overlapping strings reduce the net-color charge of the strings. To obtain $F(\xi)$, we fit the following function to the soft part of the $p_{\rm T}$ spectra with the $p_{\rm T}$ range 0.15-1.0 GeV/c of pp, Xe-Xe and Pb-Pb collisions systems at the LHC energies.

\begin{eqnarray}
	\frac{d^{2}N_{\rm ch}}{dp_{T}^{2}} =\frac{a}{(p_{0}\sqrt {{F(\xi)_{pp}/F(\xi)}^{mod}}+{p_{T}})^{\alpha}},
	\label{fitpower2}
\end{eqnarray}

where, $F(\xi)^{mod}$ is the modified color suppression factor and is used in extracting $F(\xi)$  both in $pp$ and AA collisions. The initial temperature of the percolation cluster can be defined in terms of $F(\xi)$ as \cite{Phyreport},
\begin{equation}
\label{eq7}
T(\xi) = \sqrt{\frac{\langle p^{2}_{T}\rangle_{1}}{2F(\xi)}}.
\end{equation}
By using $T_{c} = 167.7\pm2.8$ MeV \cite{bec1}  and $\xi_{\rm c} \sim 1.2$, we get $\sqrt{\langle p^{2}_{\rm T}\rangle_{1}} = 207.2\pm3.3$~MeV, from which one can get the single string-squared average momentum, $\langle p^{2}_{\rm T}\rangle_{1}$. Using this value in Eq. \ref{eq7}, we can get the initial temperature for different $F({\xi})$ values.

$\hat{q}$, and the shear viscosity-to-entropy density ratio ($\eta/s$) are transport parameters which describe the exchange of energy and momentum between fast partons and medium, are directly related to each other as~\cite{Majumder:2007}
\begin{eqnarray}
	\frac{\eta}{s} \approx\frac{3}{2}\frac{T^{3}}{\hat{q}}
	\label{etaByS_qhat}
\end{eqnarray}
Within the CSPM approach, the shear viscosity-to-entropy density ratio, $\eta/s$, can be expressed as,
\begin{eqnarray}
	\frac{\eta}{s} =\frac{T L}{5(1 - e^{-\xi})},
	\label{etaByS}
\end{eqnarray}
where, $L$ is the longitudinal extension of the string $\sim$ 1 fm~\cite{Phyreport}. 
Thus, one can get final expression for jet transport coefficient from Eq. \ref{etaByS_qhat} as:

\begin{eqnarray}
	\hat{q}\approx\frac{3}{2}\frac{T^{3}}{\eta/s}\approx\frac{15}{2}\frac{T^{2} (1 - e^{-\xi})}{L} .
	\label{eq:qhat}
\end{eqnarray}

\section{Results and discussions}

\begin{figure*}[ht!]
\centering
\includegraphics[scale = 0.35]{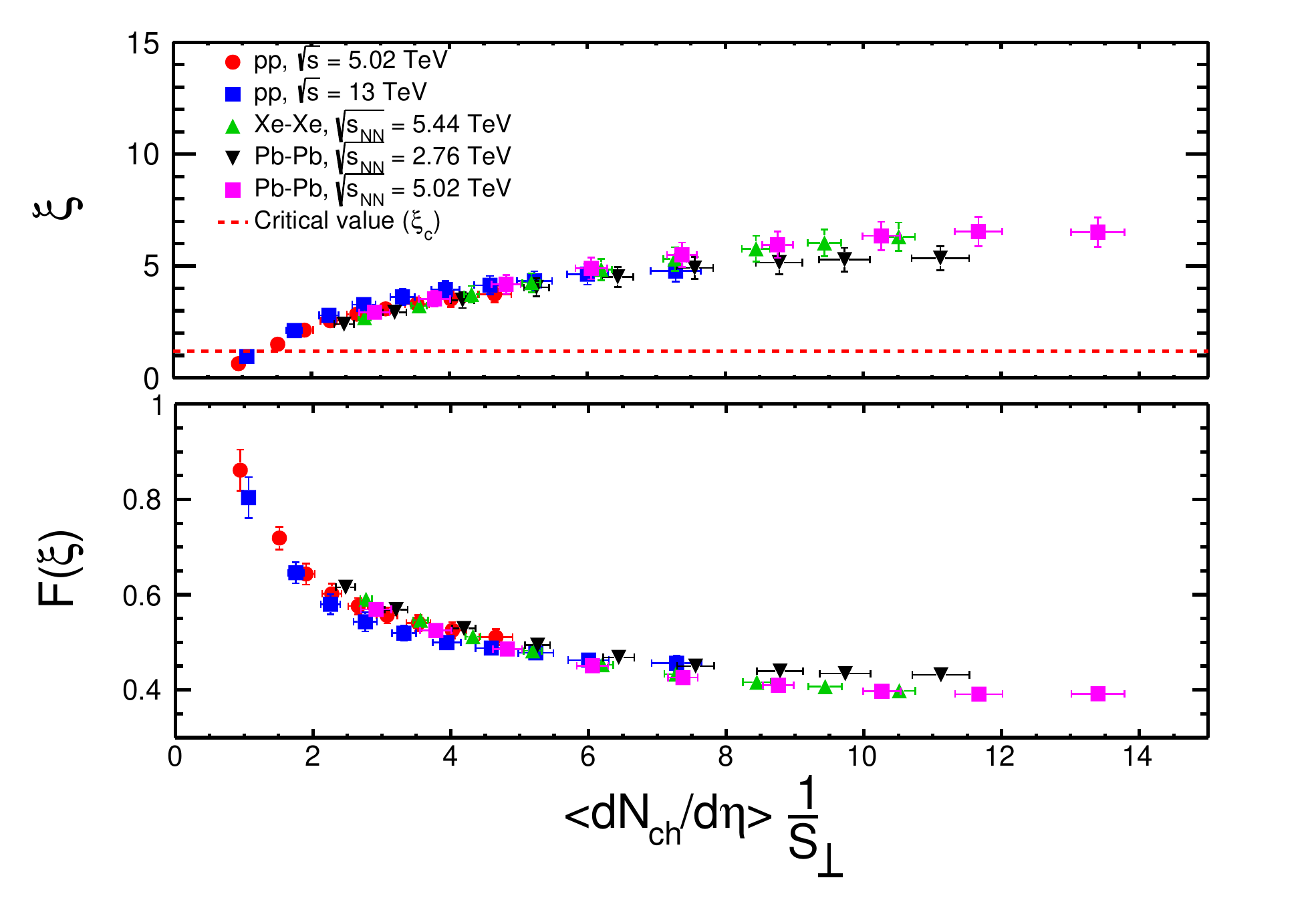}
\includegraphics[scale = 0.35]{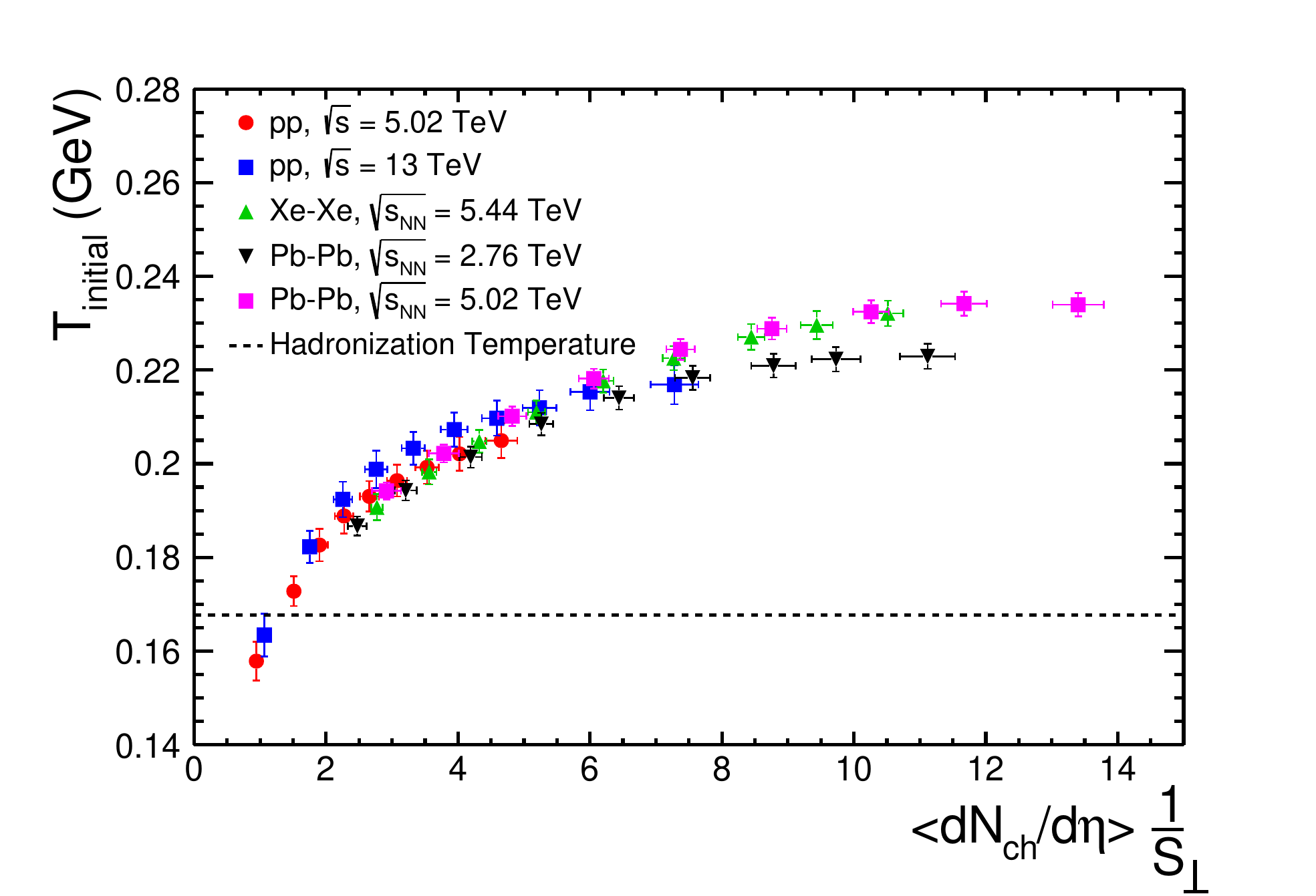}
\caption{(Color Online) Color suppression factor ($F(\xi)$) and percolation density parameter ($\xi$) (left panel) and initial percolation temperature (right panel) as functions of charged particle multiplicity density normalized by nuclear overlap area (for $pp$ collisions at $\sqrt{s}$ = 5.02 and 13 TeV, Xe-Xe collisions at $\sqrt{s_{\rm NN}}$ = 5.44 TeV, Pb-Pb collisions at $\sqrt{s_{\rm NN}}$ = 2.76 and 5.02 TeV~\cite{alice1, alice2, alice3}.}
\label{fig1}
\end{figure*}

We have plotted $\xi$ and $F(\xi)$ as functions of final charged-particle multiplicity for various collisions scaled with the nuclear overlap area in the left panel of Fig.~\ref{fig1}. For $pp$ collisions, multiplicity dependent $S_{\perp}$ is calculated using the IP-Glasma model~\cite{cross} and in case of heavy-ion collisions, the transverse overlap area is obtained using the Glauber model calculations~\cite{glauber}. We observe that $F(\xi)$ falls onto a universal scaling curve for all collisions, suggesting that the color suppression factor is independent of collision energies and collision systems in the domain of high final state multiplicity. The right panel of Fig.~\ref{fig1} shows the initial temperature estimated from CSPM as a function of $ \langle d\rm N_{\rm ch}/d\eta \rangle$ scaled by $S_{\perp}$. Here also we see a universal trend for initial temperature for all collision systems. We see that the temperature for higher multiplicity classes in $pp$ collisions is higher than the hadronization temperature, suggesting a deconfied medium formation.

\begin{figure*}[ht!]
\centering
\includegraphics[scale = 0.35]{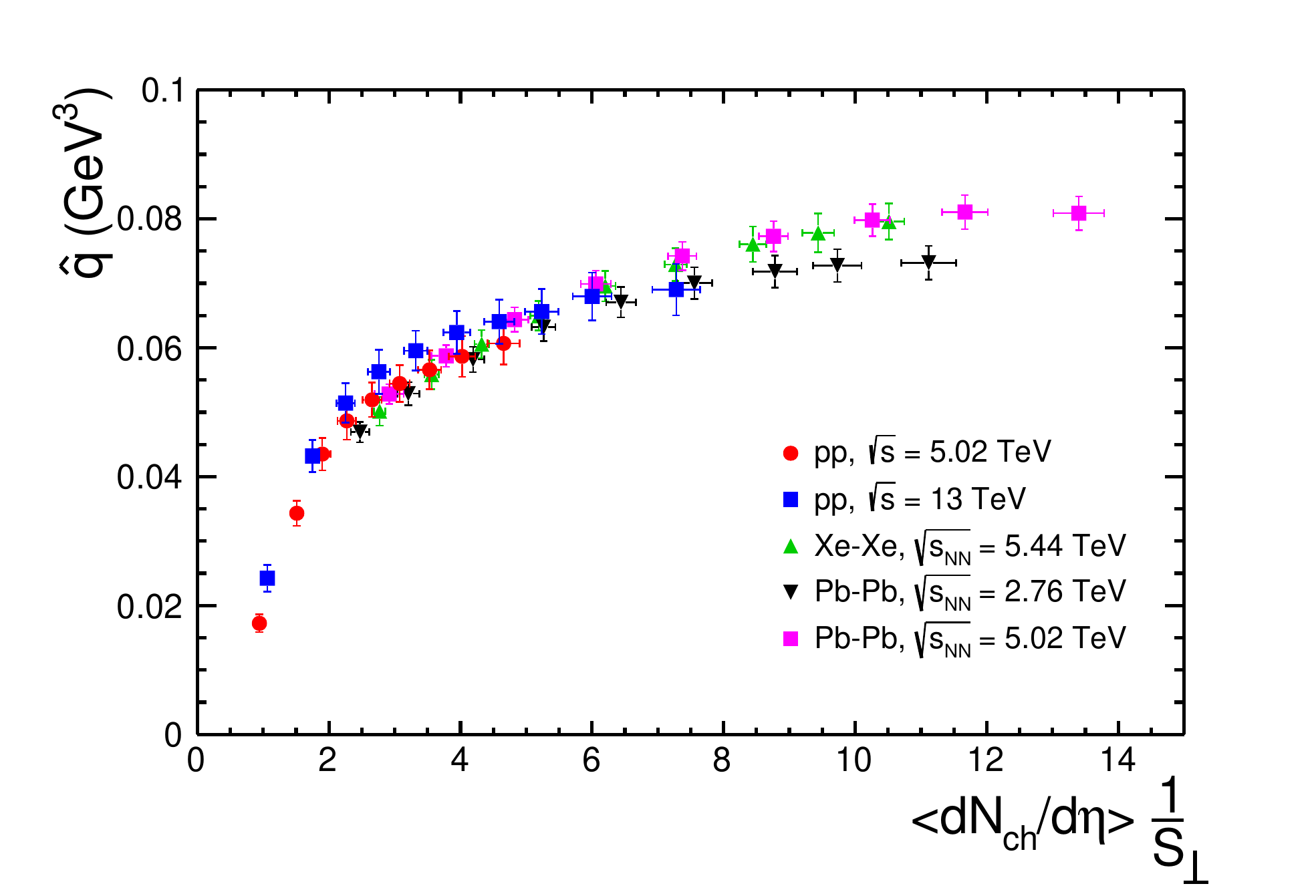}
\includegraphics[scale = 0.35]{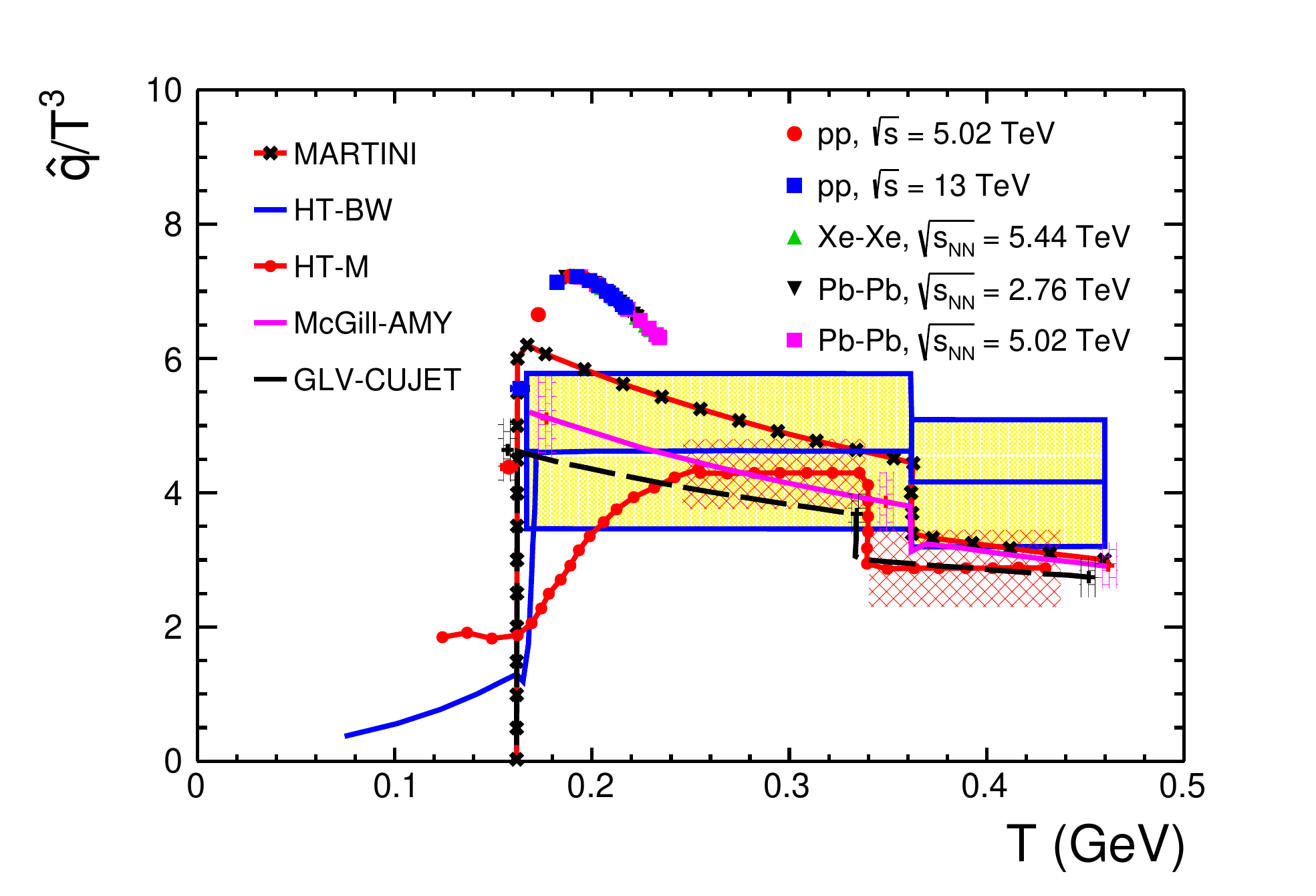}
\caption{(Color Online) Jet transport coefficient as a function of charged particle multiplicity density normalized by nuclear overlap area (left panel) and scaled jet transport coefficient as a function of initial percolation temperature (right panel) (for $pp$ collisions at $\sqrt{s}$ = 5.02 and 13 TeV, Xe-Xe collisions at $\sqrt{s_{\rm NN}}$ = 5.44 TeV, Pb-Pb collisions at $\sqrt{s_{\rm NN}}$ = 2.76 and 5.02 TeV~\cite{alice1, alice2, alice3}.}
\label{fig2}
\end{figure*}

On the left panel of Fig.~\ref{fig2}, we have plotted $\hat q$ as a function of charged particle multiplicity scaled with transverse overlap area for various collision systems. We observe a steep increase in the trend at lower charged particle multiplicities in $pp$ collisions, which gets saturated at very high multiplicity. This behavior suggests that at lower multiplicities, the system is not dense enough to quench the partonic jets highly. However, with the increase in multiplicity, the quenching of jets becomes more prominent. On the right panel of Fig.~\ref{fig2}, we have plotted $\hat q/T^{3}$ as a function of initial temperature. For comparison, we have also plotted the results obtained by the JET collaboration using five different theoretical models that incorporate particle energy loss in the medium. We observe that $\hat q/T^{3}$ obtained from the CSPM approach has a similar kind of behavior as observed by JET collaboration. 

\section{Summary}
In summary, we have estimated the jet transport coefficient using CSPM for various systems at LHC energies. Our results from CSPM show a similar kind of trend as the results obtained from the JET collaboration. We also observe critical behavior at the hadronization temperature in the scaled $\hat q$ plot suggesting a change in the dynamics of the system.

\end{document}